\documentclass[structabstract]{aa}  
\usepackage{graphicx}
\usepackage{longtable}
\usepackage{txfonts}
\usepackage{textcomp}
\usepackage {ulem}
\usepackage[]{natbib}

\begin{document}
\title{Multi-object spectroscopy of stars in the CoRoT fields I
    \subtitle{Early-type stars in the CoRoT-fields IRa01, LRa01, LRa02}
\thanks{based 
         on observations obtained Anglo-Australian Telescope in program
         07B/040 and 08B/003.}}
\authorrunning{Sebastian et al.}
\titlerunning{Stars in the CoRoT fields I}
\author{D. Sebastian\inst{1},
        E.W. Guenther\inst{1},
        V. Schaffenroth\inst{2},
        D. Gandolfi\inst{3},
        S. Geier\inst{2},
        U. Heber\inst{2},
        M. Deleuil \inst{4},
        C. Moutou \inst{4}
          }
   \institute{Th\"uringer Landessternwarte Tautenburg,
              Sternwarte 5, D-07778 Tautenburg, Germany\\
              \email{sebastian@tls-tautenburg.de}
         \and
              Dr. Karl Remeis-Observatory \& ECAP, Astronomical Institute, 
	      Friedrich-Alexander University Erlangen-Nuremberg, Sternwartstr.~7, D 
	      96049 Bamberg, Germany
         \and
               Research and Scientific Support Department, 
               European Space Agency, Keplerlaan 1, 
               2200AG, Noordwijk, The Netherlands
         \and
              Laboratoire d'Astrophysique de Marseille, 
              38 rue Fr\'ed\'eric  Joliot-Curie, 
              13388 Marseille Cedex 13, France 
              }

   \date{Received July 10, 2011; accepted July 25, 2011}

  \abstract 
{Observations of giant stars indicate that the frequency of giant planets
  is much higher for intermediate-mass stars than for solar-like stars. 
  Up to now all known planets of giant stars orbit at relatively large distances from their host stars.
  It is not known whether intermediate-mass stars also had a large number of close-in planets when 
  they were on the main sequence, which were then engulfed when the star became a giant star. 
  In order to understand 
  the  formation and evolution of planets it is thus important to find out whether main-sequence stars 
  of intermediate-mass have close-in planets or not.}
{A survey for transiting planets of intermediate-mass stars would be ideal
 to solve this question, because the detection of transiting planets is not affected
 by the rapid rotation of these stars. With CoRoT it is possible to detect 
 transiting  planets around stars up to a spectral type B4V. As a first step for an efficient survey we need to identify intermediate-mass stars in the CoRoT-fields which than can be used as an input list.}
{In order to compile the input list we derive the spectral types of essentially all O, B and A stars 
down to 14.5 mag in the CoRoT fields IRa01, LRa01, LRa02 taken with the multi-object spectrograph AAOmega. 
The determinaton of the spectral types will be done by comparing the spectra with template spectra 
from a library.
}
{In total we identify 1856 A and B stars that have been observed with
  CoRoT.  Using multiple spectra of the same stars, we find that
  the accuracy of the resulting spectral classification is $1.61\pm0.14$ sub-classes 
  for A and B stars.}
{Given the number of planets that have been detected in these fields
  amongst late-type stars, we estimate that there are one to
  four transiting planets of intermediate-mass stars waiting to be
  discovered.  Our survey  not only allows to carry out a dedicated planet search program
  but is also essential for any kind of studies of the light curves of early-type stars in the CoRoT
  database. We also show that it would be possible to extend the survey to
  all fields that CoRoT has observed using photometrically determined
  spectral types.
} 
 \maketitle
%

\section{Introduction}
\subsection{The frequency of planets orbiting A-type stars}

Now more than 400 planets orbiting low-mass main-sequence stars (G-type stars and later) have
been detected indirectly by means of precise radial-velocity (RV)
surveys and more than 100 transiting planets have also been
found (The Extrasolar Planets Encyclopaedia\footnote{http://www.exoplanet.eu, 2 August 2011}). 
This data gives us detailed information about the properties of
planets orbiting main-sequence stars with $M\leq
1.2\,M_{\odot}$. However, our knowledge about the planets orbiting
intermediate-mass stars (A and B-type stars) is still very limited. 
The reason is that it is difficult to detect planets of main-sequence stars
with masses larger than ($1.2\,M_{\odot}$) using the radial-velocity (RV)
technique. This is because most of these stars rotate rapidly which reduces the 
accuracy of the RV-measurements significantly due to the broadening of the lines  

The most important source of information about planets of
intermediate-mass stars are RV-surveys of giant stars.
About 30 planets orbiting giant stars have been discovered, many of
them orbiting stars that were more massive than the Sun when they were
on the main sequence.  All groups studying these stars come to the
same conclusion: intermediate-mass stars have a higher frequency of
planets than solar-like stars (G-type stars). For example, \cite{johnson10a,johnson10b}
 conclude that the frequency of planets
orbiting stars in the mass range between 1.5 and 3 $M_{\odot}$ is a
factor of 2 to 4 higher than for stars of 1.0 $M_{\odot}$, in good agreement
with the theoretical expectations \citep{kennedy08}.

The large frequency of planets orbiting giant stars is particularly
surprising, because all known planets of giant stars have
long orbital periods (semi-major axis between 0.5 and 2.6 AU),
and are relatively massive with a median of 5.3 $M_{\textrm{\small{Jup}}}$ (The Extrasolar Planets Encyclopaedia). 
It is interesting to note that the lack of close-in planets orbiting giant stars 
is in relative terms even more extreme than for solar-like stars. Planets of 
giant stars have distances $>8\,R{_*}$, whereas 
planets of solar-like stars have distances that can be much closer than 
$3\,R_{*}$. As showed by \cite{bowler10}, the
differences between the properties of planets orbiting giant and
low-mass main-sequence stars is not an artifact of the sample size 
but the two populations are different. 
\cite{alibert11} modeled from planet-formation theories the 
properties of the planetary population for different stellar masses.  
They predict that stars with 2 $M_{\odot}$ should have massive, close-in planets, 
if the disc-mass does depend on the stellar mass. If it doesn't, 
they do not expect close-in planets. Moreover they predict that the properties 
of close-in planets orbiting intermediate-mass stars depend much stronger on 
the properties of the disc than of those that orbit solar-like stars.

The crucial question thus is, what caused the observed lack of
planets orbiting giant stars within 0.5 AU.
There are three possible scenarios. The first hypothesis is
that the lifetime of the gaseous disk of intermediate-mass stars is
so short that there is not enough time for the planets to
migrate inwards \citep{currie09}.  A second hypothesis is
that the planets migrate outwards during the evolution
of the star. It is in fact true that planets move outwards due to the 
mass-loss of the star, but this effect is not large enough to explain 
the complete lack of planets within 0.5 AU. A third possibility is that 
intermediate-mass stars have close-in planets when they
are on the main sequence but these are engulfed when the star
becomes a giant star \citep{sato08}.
Calculations by \cite{villaver09} show that tidal
interaction can lead to the engulfment of close-in planets 
by evolved stars. Because the engulfment of planets may effect the 
evolution of stars, it would be important to find out whether this process
is common, or rare. Which of the three possibilities is correct can only be 
solved by a dedicated survey for close-in planets orbiting main-sequence 
stars of intermediate-mass.

There are actually a few planets known that orbiting intermediate-mass stars before they become 
giant stars. Using direct  imaging techniques, planets of the A stars  $\beta$
Pic \citep{lagrange09}, HR 8799 \citep{marois08} and Fomalhaut have been detected
\citep{kalas08}. Since the time for the formation of the core 
scales with the orbital period and all of these planets have orbital periods of 
more than 15 years, the very existence of these planets tells us that there
must have been also enough time for close-in planets to form. However for proving 
the predictions for close-in planets these long periodic planets are not suitable.  

\cite{johnson10b} reported the discovery of a giant planet
orbiting HD 102956 at a close distance of 0.081 AU. The planet was 
detected by precise RV-measurements of the 1.68 $M_{\odot}$- star which is classified 
as a A-sub-giant star.

The only known transiting planet orbiting an A-type star is WASP-33b. 
It is a very close-in planet with a orbital distance of 0.026 AU \citep{cameron10}.
It is interesting to note, that this planet is currently the hottest planet known 
\citep{smith11}. Studying such planets helps us to understand whether 
evaporation plays an important role in the evolution of planets. 
Because the amount of evaporation depends on the density of the planet and the amount 
of radiation that the planet receives \citep{lammer09}, 
we need a sample of close-in planets were these parameters are determined. 
Another interesting aspect is that \cite{herrero11} detected $\delta$ 
Sct -like oscillations that might be induced by the planet. If all close-in planets of A stars
would induce oscillations, it would be easy to understand why just one
has been found. It is also interesting to note that there are transit candidates of
early-type stars published by the CoRoT \citep{carpano09} and Kepler 
space-missions \citep{borucki11}. Since the spectral types of these candidates are 
not well known, none of these candidates were confirmed as planet of an intermediate-mass star 
yet. These discoveries show the importance of a dedicated search program to detect planets 
orbiting A-type stars.


\subsection {A new survey for transiting planets of A and B stars}

As outlined above, transit search programs are ideal for finding close-in planets of early-type stars. Space-based transit search 
programs have a number of advantages over ground-based search programs: 
not only is the photometric accuracy obtained much higher but the excellent 
time-coverage  allows to remove the oscillations more efficiently. 
The removal of stellar oscillations would be particularly important 
if it turns out that close-in planets can induce $\delta$ Sct -like oscillations.
Last not least,  a long time-series allows to average many
transits, which makes the detection of shallow transits easier.

\begin{figure}[h!]
\includegraphics[height=.26\textheight,angle=0]{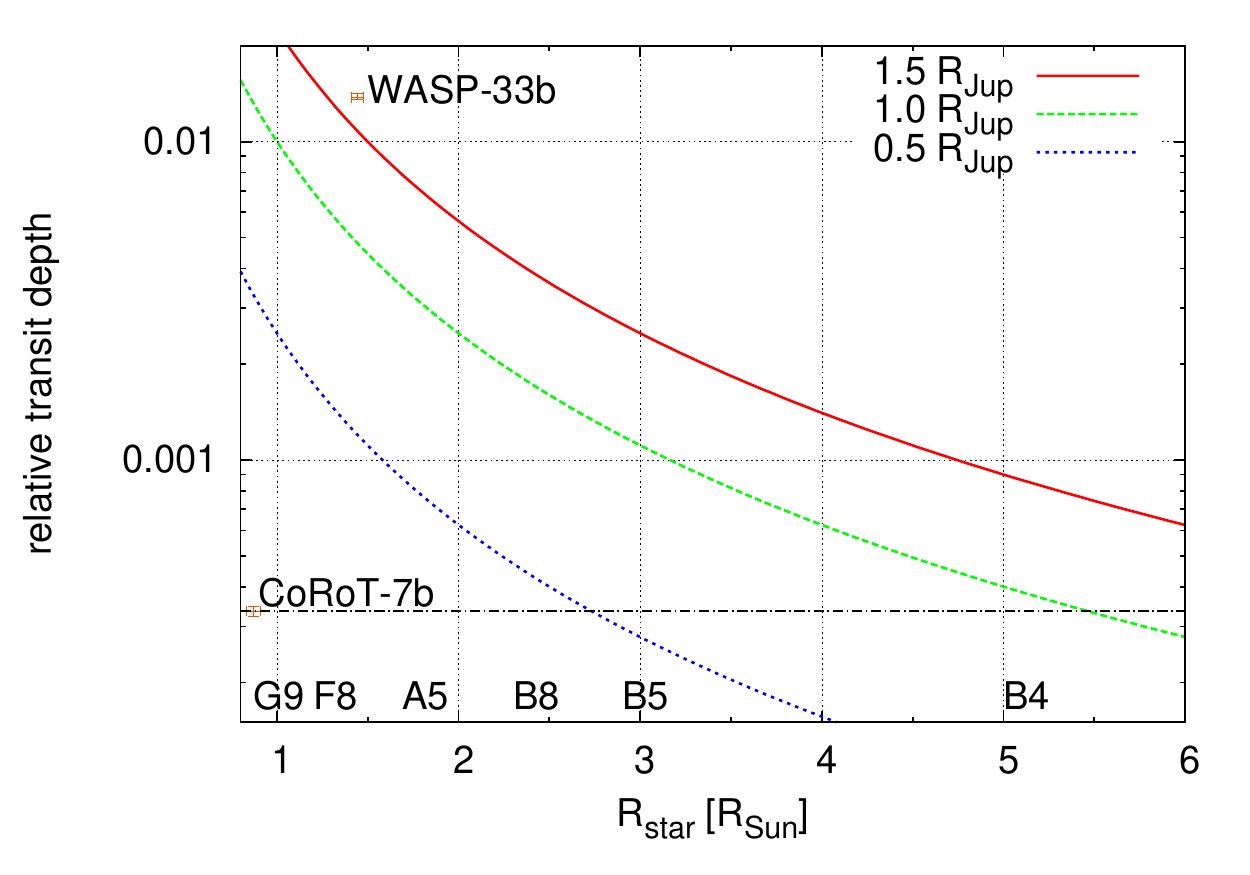}
\caption{Detectability of planets of 1.5, 1.0 and 0.5 $R_{\textrm{\tiny{Jup}}}$ orbiting
stars of different spectral types.  With ground-based
telescopes it is possible to detect transits with a depth of about 1\%, whereas
CoRoT has already detected a planet with a transit-depth of 0.034\%. This means 
that CoRoT has the capability of detecting planets orbiting stars up to B4V.}
  \label{LimitsRadiusPlanet}
\end{figure}

We intend to use data obtained with the CoRoT (COnvection ROtation \& planetary
Transits) satellite \citep{auvergne09}. CoRoT has detected a transiting planet with
a relative transit depth of $3.4\cdot10^{-4}$ orbiting a star of 11.7 mag
\citep{leger09}. This means that CoRoT has the capability to detect
a transiting planet with the size of Jupiter orbiting a B4V star
(Fig.\ref{LimitsRadiusPlanet}). It might be possible to detect planets of A stars 
down to 14.5 mag. This compares favorably with ground-based
observations which achieve an accuracy of only about 0.01. 
In the case of the long runs, the fields were observed continously by CoRoT for up to 150 days. In the normal mode, the time-sampling 
is 8.5 minutes. 

For such a transit search program, we have to identify suitable targets
first, and then carry out a dedicated analysis of the light curves. 
We selected the first three fields that CoRoT observed in
the so-called anti-center "eye": IRa01 (RA $100.9^{\circ}$ to 
$102.6^{\circ}$, DEC $-0.5^{\circ}$ to $-3.3^{\circ}$), 
LRa01 (RA $100.1^{\circ}$ to $101.5^{\circ}$, 
DEC $1.1^{\circ}$ to $-1.6^{\circ}$), 
and LRa02 (RA $101.9^{\circ}$ to $103.5^{\circ}$, 
DEC $-3.0^{\circ}$ to $-5.9^{\circ}$).  
Prior to the launch of the satellite, these fields
were observed in B, V, R, and I \citep{deleuil06}.  
These observations were cross-correlated with 2MASS J, H, K$_{\textit{s}}$ magnitudes 
\citep{cutri03} and used to perform a dwarf-giant
discrimination and spectral classification. The photometric measurements
in both sets of filters and the spectral classification are available through 
the database EXODAT\footnote{The EXODAT database is continuously updated and online available: http://cesam.oamp.fr/exodat/} \citep{deleuil09}

However, we did not apply any pre-selection criteria but to analyze all stars  down to $M_V=14.5$ 
spectroscopically. The observation and spectroscopic analysis of all stars on the one hand provides us with a 
suitable target list and on the other hand gives the opportunity to verify the accuracy, if we 
would use a target list compiled from photometric data (see Section \ref{phot_compare}).

We report a survey for spectral classification of stars in the CoRoT-fields IRa01, LRa01, and LRa02 
and divide it into two parts. One containing  a catalogue of all O, B, and A stars found in this survey. This
catalogue is the input catalogue for a dedicated search program for 
planets of intermediate-mass stars. The second part of the survey
contains the catalogue of F,G,K and M stars \citep[paper II]{guenther11}
which aims in comparing statistically the
spectral types of planet host stars with the sample of stars that has
been observed. We exclude the early-type stars from paper II, because 
no dedicated survey of planets of early-type stars has been carried out so far. 

A survey for early-type stars in the CoRoT-fields is not only
important for planet search programs but also essential for any study
of the CoRoT-light curves of early-type stars. 

For example are the fields of most of the surveys for hot-subdwarf stars 
(sdO/B), which are of special interest for close binary research and 
asteroseismology (for a review, see \citealt{heber09}), arranged in a 
perpendicular direction to the Galactic plane. The spectroscopic identification of O and B-type 
stars is the first step for the identification of the yet unknown population 
of hot-subdwarf stars situated in the Galatic plane. The analysis of 
high-precision sdB light curves obtained by the Kepler mission \citep[e.g.][]{ostensen10} led to 
very interesting results and show us, that we can learn a lot about 
the properties of such objects by extending these analysis to the CoRoT-felds. 

Another application of our survey are studies of rapidly oscillating Ap (roAp), or 
$\gamma$ Doradus stars. Such objects have been identified in the Kepler 
field \citep[e.g.][]{balona11} but not yet in the CoRoT-fields.

Using the multi-object spectrograph FLAMES/GIRAFFE \cite{gazzano10} has studied the stellar population in the CoRoT-fields LRa01, 
LRc01, and SRc01, but since there is only one A-star in that survey which was
also observed by us (CoRoT 102677302), we compare their results with 
our results in paper II \citep{guenther11}. 

\section{Spectroscopic observations}

As part of the CoRoT follow-up observations, we observed stars in
IRa01, LRa01, and LRa02 with the AAOmega multi-object spectrograph
mounted on the AAT (Anglo-Australian Telescope/Australian Astronomical
Telescope).  The AAOmega multi-object-spectrograph is ideal for this
purpose, because it has a field of view of 2 degrees, which matches
perfectly to the size of the CoRoT fields which have a size of 1.8x3.6 
square-degrees \citep{saunders04,smith04}.

The data was obtained in two campaigns. The first campaign was carried
out in the period from the 13 to the 20 January 2008.  Unfortunately, due to the 
weather conditions, observations could only be carried out in the first two nights. 
The second campaign was carried out from the 28 December 2008 to the 4 January
2009 in which observations were carried out in all eight nights.

We used the AAT software tool ``Configure'' for our target allocation 
in order to find the optimum configuration of the fibres. In each setting we 
typically placed 350 fibres onto target stars, and 25 onto the sky
background. We also added stars that are not observed by CoRoT 
to maximize the number of stars observed in each setting.
In order to optimize the exposure time and in order to avoid cross-talk 
between the fibres, all stars observed in each setting have in general 
within half a magnitude the same brightness. We started our observations 
with the fields containing the brightest stars and subsequently used settings 
for fainter and fainter stars. 

\begin{figure}[h!]
  \includegraphics[height=0.25\textheight]{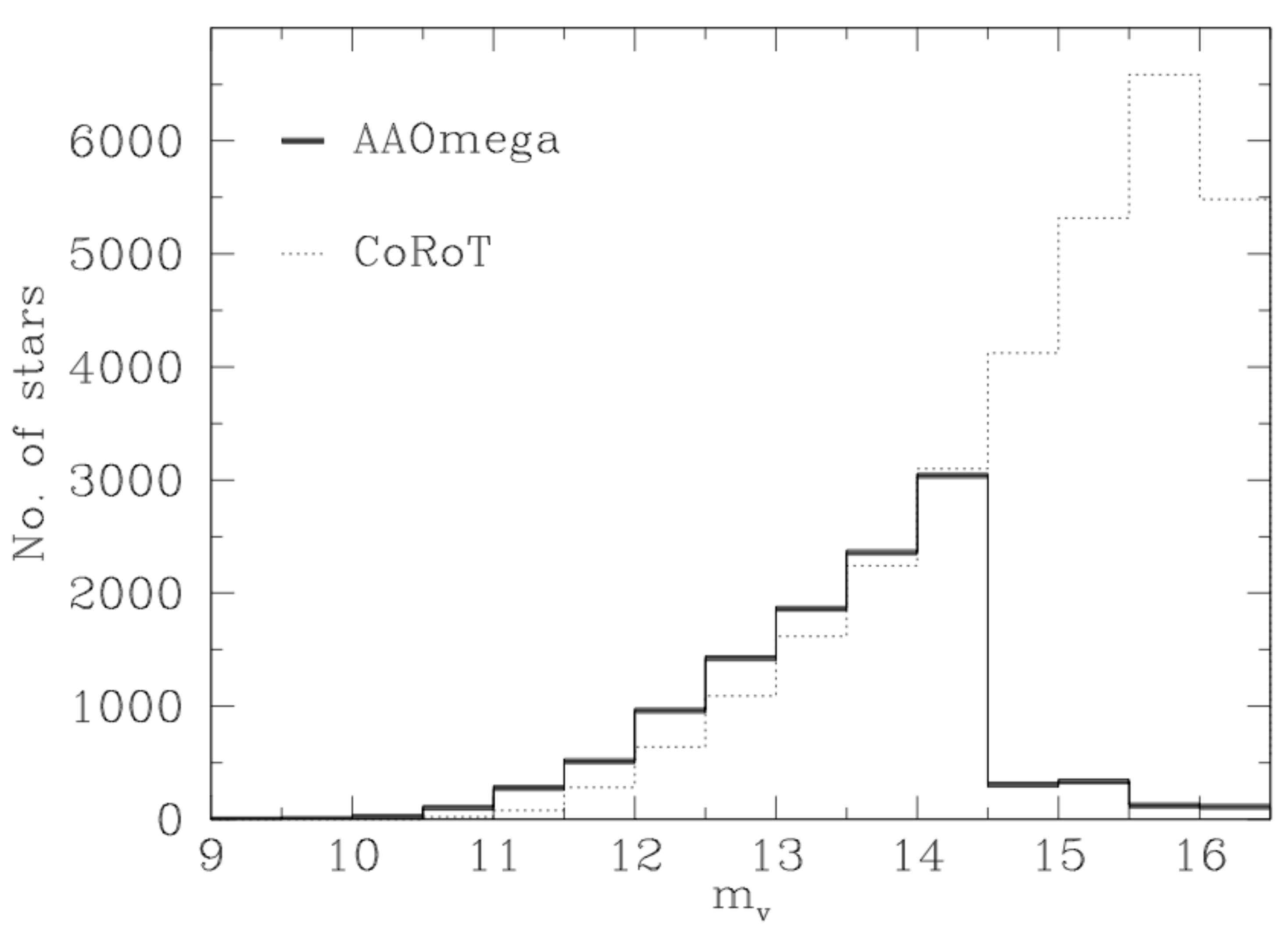}
\caption{Number of stars observed by CoRoT (dotted line) 
              compared to the number of stars which we analyzed
               for each brightness interval (solid line).}
  \label{histMag}
\end{figure}

Fig.\ref{histMag} shows a histogram of the 31\,405 stars observed by CoRoT in 
the three fields and the 11\,466 stars observed and analyzed in this work. 
Our sample is essentialy complete down to 14.5 mag in V.  This is roughly 
the limit for which CoRoT obtains light curves in three colours, fainter stars are
only observed in the monochromatic mode, and it is also roughly the limit
to which we can detect planets of A stars. 

Guiding was done by placing a fibre bundle onto a bright star in the field.
The field rotation was monitored by placing additional fibres onto 
typical 6 stars close to the edge the field-of-view.  We 
used the 580V grating in the blue arm and the 385R in the red arm.
The spectra cover the range from 3740 to 5810 \AA \, in the
blue arm, and 5650 to 8770 \AA \, in the red arm. The resolution is
about R=1300 for both arms. Each field was observed for 30 to 45
minutes. In order to avoid saturation, and in order to make the
removal of cosmic rays easier, we split the observing time spend on
each field into three or more exposures.

All calibration frames like flats, bias-frames and arcs (using a CuAr+FeAr-lamp 
for wavelength-calibration) were taken in the usual way. Bias frames were taken each day in the 
afternoon with the dome closed and the light switched off. Since the detector is operating 
at a temperature of 160 K \citep{sharp06} no dark frames were required. 
Flats and arcs were taken during the night before the observation 
of each field. We subtracted the sky-background from each spectrum using the
average spectrum of the night-sky taken with the sky-fibres, taking
the throughput of each individual fibre into account. 
We measured the throughput of each individual fibre 
for each setting always after the fibres were configured,
because the  throughput changes depending of how the fibre is 
bended.  As usual, the fibres for the first two fields were positioned  
in the afternoon. We then measured the throughput for each fibre
by obtaining spectra during the evening dawn. The same procedure 
was also done using the dawn in the morning for the last two fields. 
Using spectra of the blank sky, we measured the throughput 
for the fields observed in the middle of the night. The sky
subtraction is not critical, because we observed only stars brighter
than 15.0 mag and the observations were carried out during dark time.
Flux-calibration was finaly done by observing 5 to 10 well-known stars in each field.

\section{Classification of the spectra}

The spectral types of the stars were determined by comparing the observed
spectra with templates from a library of spectra automaticaly.  The spectral type
of a star is then given by the best matching template 
\citep[see also][]{gandolfi08}. As library 
we used ``The Indo-U.S. Library of Coud\'{e} Feed
Stellar Spectra'' \citep{valdes04} which contains a
dense grid of spectra of different spectral types and luminosity classes.  In order to
fit an observed spectrum to a template, we iteratively adjust the
radial velocity, the extinction \citep{binney98} and the flux of the observed
spectrum by minimizing the sum of the differences between the template
and the observed spectrum squared ($\sigma^2$). 
 \begin{figure}[h!]
  \includegraphics[height=.25\textheight]{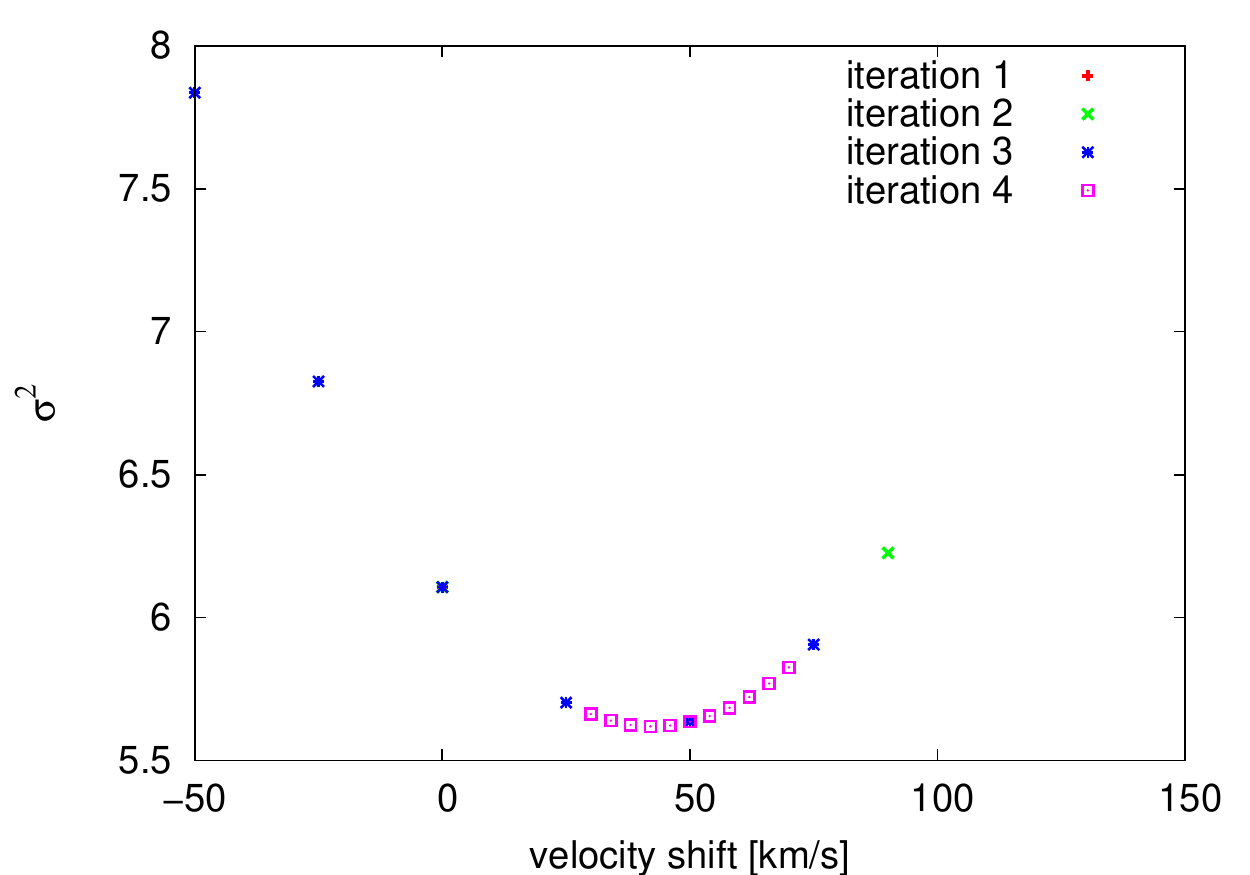}
\caption{The spectral types are obtained iteratively by fitting the
  observed spectra to templates by minimizing $\sigma^2$. Shown here
  is the derived $\sigma^2$ vs. the velocty-shift. The best
  match is obtained at the minimum of $\sigma^2$.}
  \label{shift}
\end{figure}

\begin{figure}[h!]
  \includegraphics[height=.25\textheight]{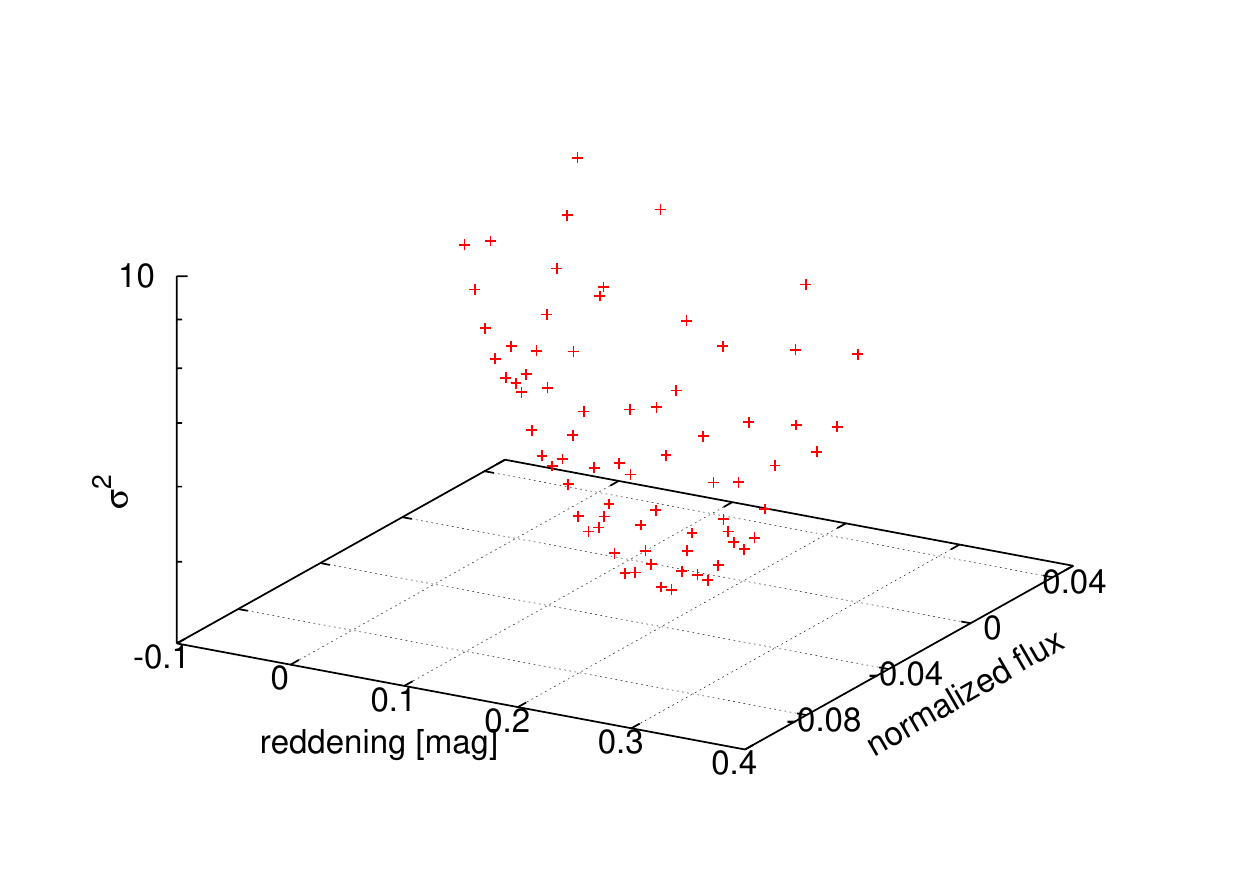}
\caption{Same as Fig.\ref{shift} but for the flux and the reddening.}
  \label{box}
\end{figure}
\begin{figure}[h!]
  \includegraphics[height=0.25\textheight]{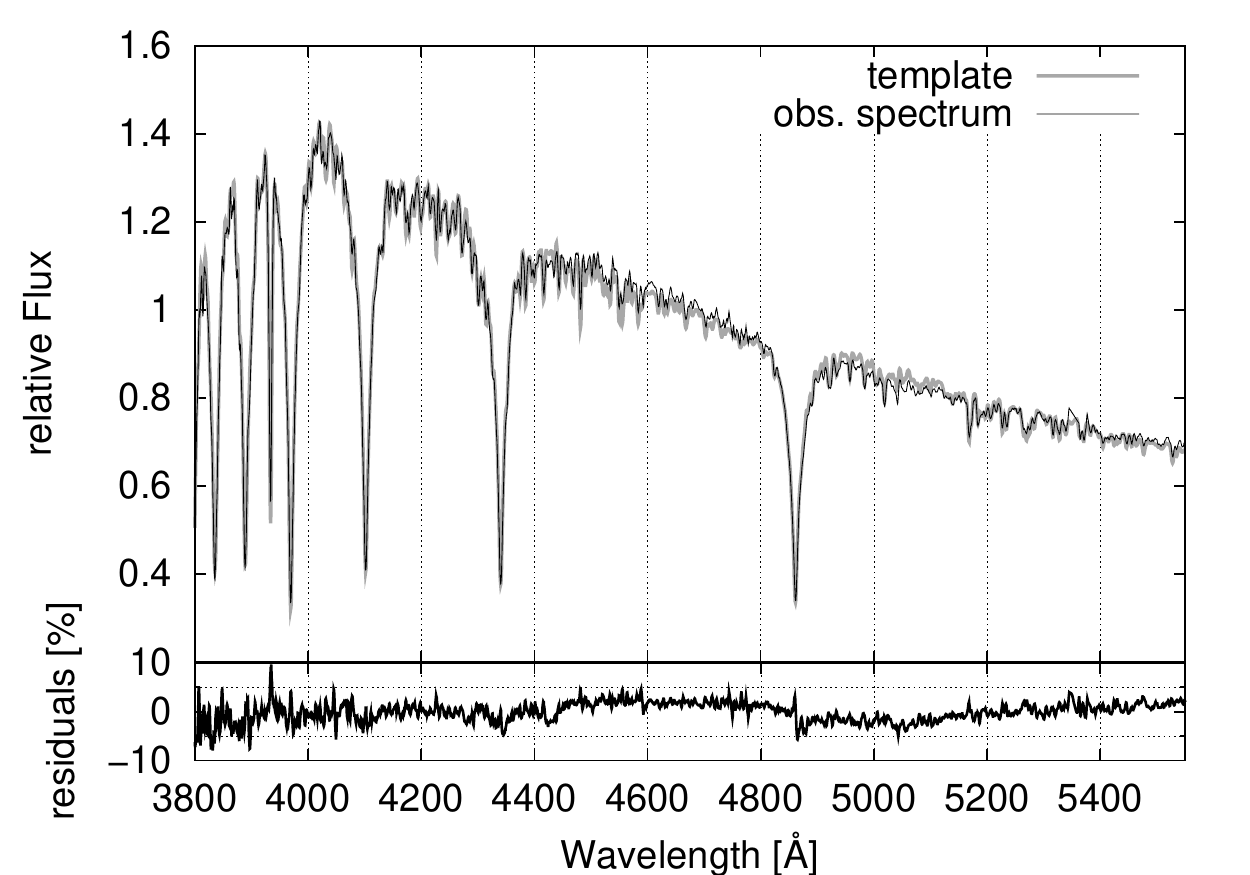}
\caption{Spectrum of an A5V-star (black line), together with the
         template (grey line). The lower panel shows the residuals between the two spectra.}
  \label{spec}
\end{figure}

\begin{figure}
 \includegraphics[height=.25\textheight]{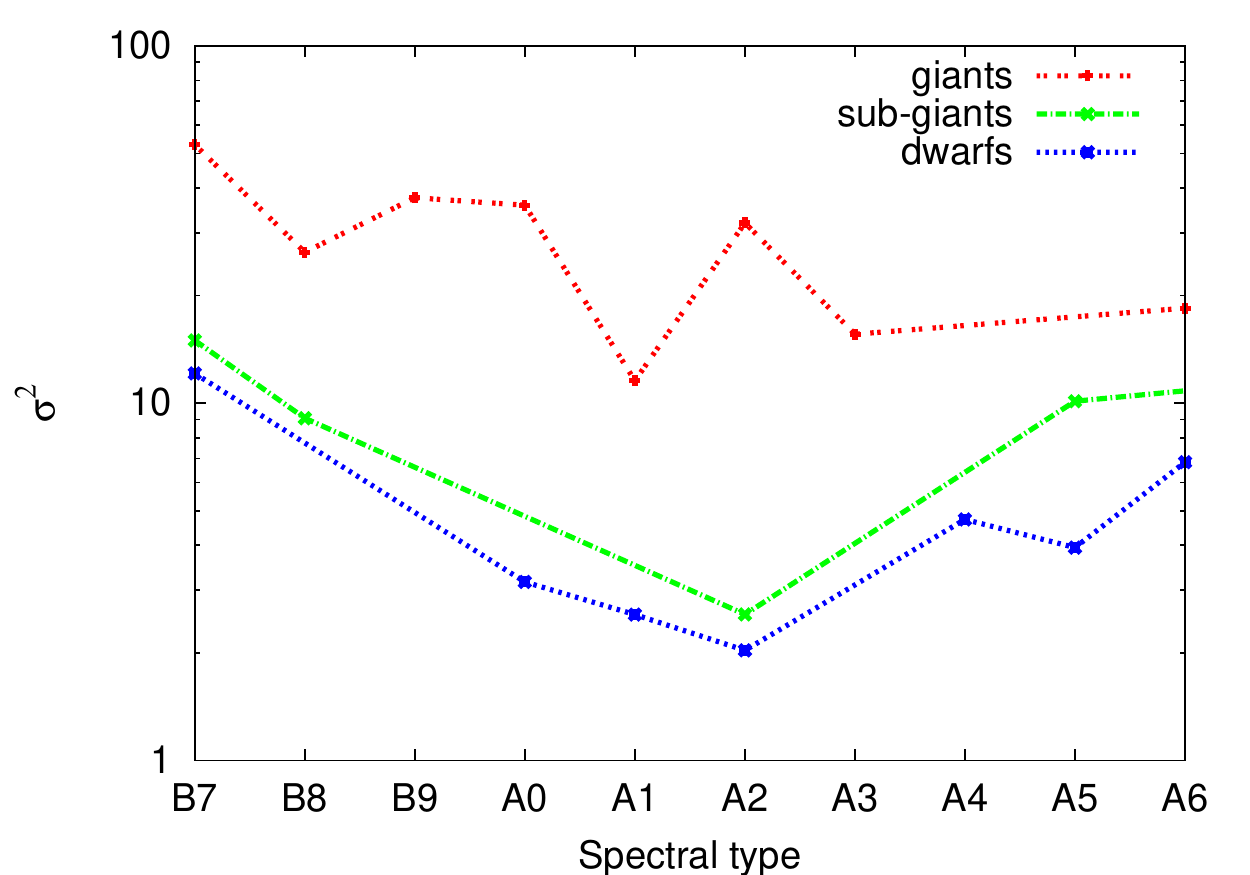}
  \caption{Shown is $\min (\sigma^2)$ for templates of different
   spectral types and luminosity classes.  $\min \sigma^2$ is
   smallest for the A2V template. }
  \label{lumi}
 \end{figure}

 
Figs. \ref{shift} and \ref{box} show how the
radial-velocity, the flux and the extinction (reddening) were determined by minimizing
$\sigma^2$. For each template we thus derive $\min(\sigma^2)$.
Fig. \ref {lumi} shows how $\min(\sigma^2)$ changes for templates
of different spectral types. The best matching template is the
one with the smallest $\min(\sigma^2)$, and the spectral type
of that template is also the spectral type of the star, A2V in this case.
Fig. \ref {lumi} also shows that we can reliably distinguish between 
dwarf stars and giant stars but for A and B stars the difference between 
dwarf stars and sub-giant stars is so small that we can not distinguish between 
them. Fig. \ref{spec} shows an example how well the template matches 
an observed spectrum. The black line is the observed spectrum of the star, 
the grey line is the template spectrum.

\section{The accuracy of the method}

\begin{figure}
  \includegraphics[height=.25\textheight]{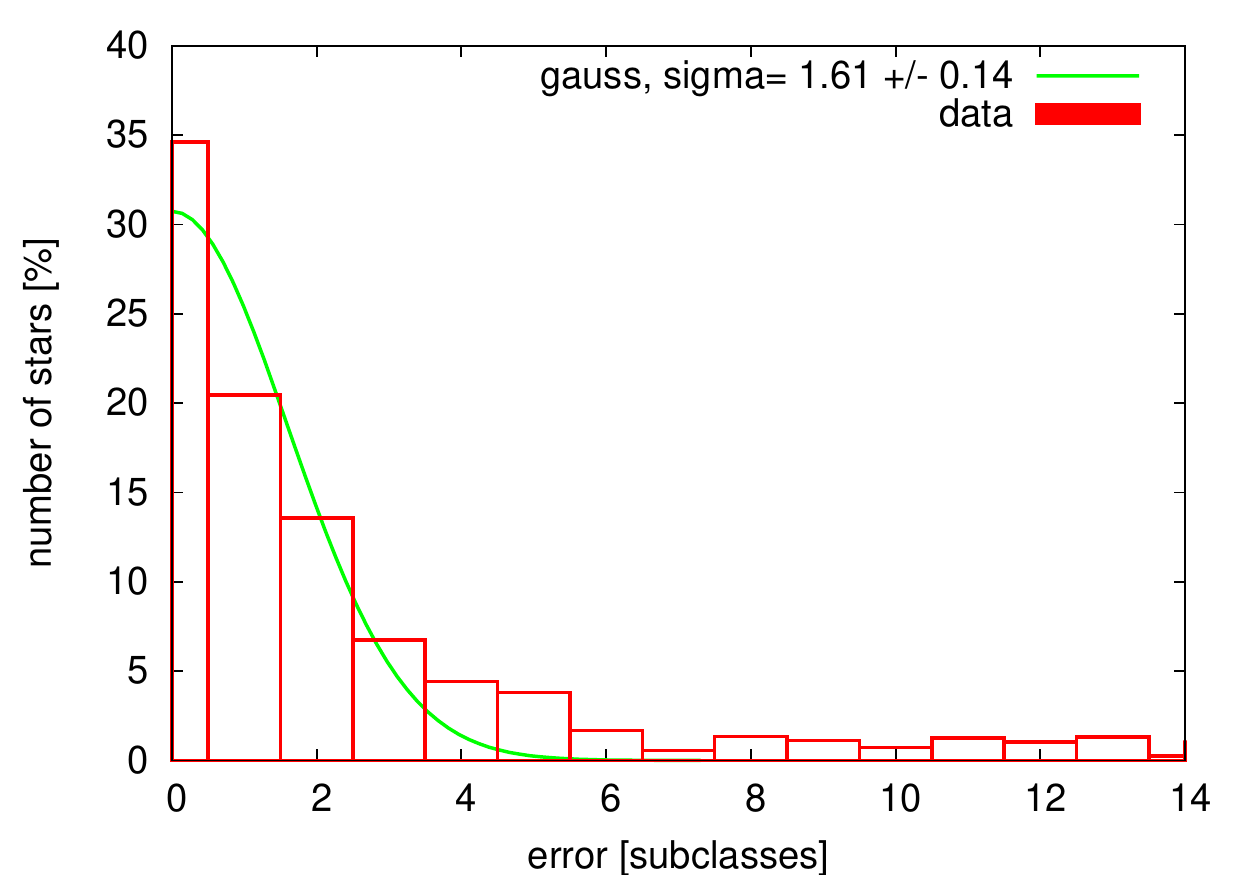}
  \caption{The error in sub-classes for A and B stars, derived 
           from the difference of the spectral types obtained for stars 
           of which several spectra were taken.}
  \label{errorSpec}
\end{figure}

Due to some overlap in the observations for  931 A and B stars we obtained two or more spectra.
By comparing the spectral types derived for the same stars from 
different spectra, we can thus determine the error of our method.
Fig.\ref{errorSpec} shows a histogram of the differences in
sub-classes between the determination of the spectral type 
for the same stars.  For A and B stars the error is on average
$1.61\pm0.14$ sub-classes. 

There are, however, few outliers
were the difference is larger than 4 sub-classes. Most likely, these stars are intrinsically 
variable or show chemical peculiarities. Outliers can also be caused by 
instrumental problems, like broken or displaced fibres which 
cause  noisy-spectra. Another instrumental
problem which affects a few spectra, is a wave-like fringe pattern. In
AAOmega starlight enters through a small rectangular prism, which then
reflects the light into the fibre (for a detailed
description of the instrument, see \citealt{saunders04,smith04}). This prism is normally glued to the
fibre entrance but in some cases, the prism is not perfectly glued to
the fibre entrance.  In such cases there is a little gap between the
fibre and the prism, the resulting ``Newtonian rings'' then cause the
wave-like pattern in these spectra, which reduces the accuracy with
which the spectral types can be determined. 

However, in all cases when we had two or more spectra of the same star
we inspected the spectra individually, and disregarded spectra
with low S/N or `Newtonian rings''. The error of our spectral classification thus
is certainly smaller than $1.61\pm0.14$ sub-classes.

Regarding the accuracy of the classification, it turns out that the spectral 
classification give slightly different results by comparing with 
libraries published by different authors \citep{borgne03,jacoby84}. For this 
reason we derive the spectral-types by comparing with templates of one 
library.

\begin{figure}[h!]
  \includegraphics[height=.25\textheight]{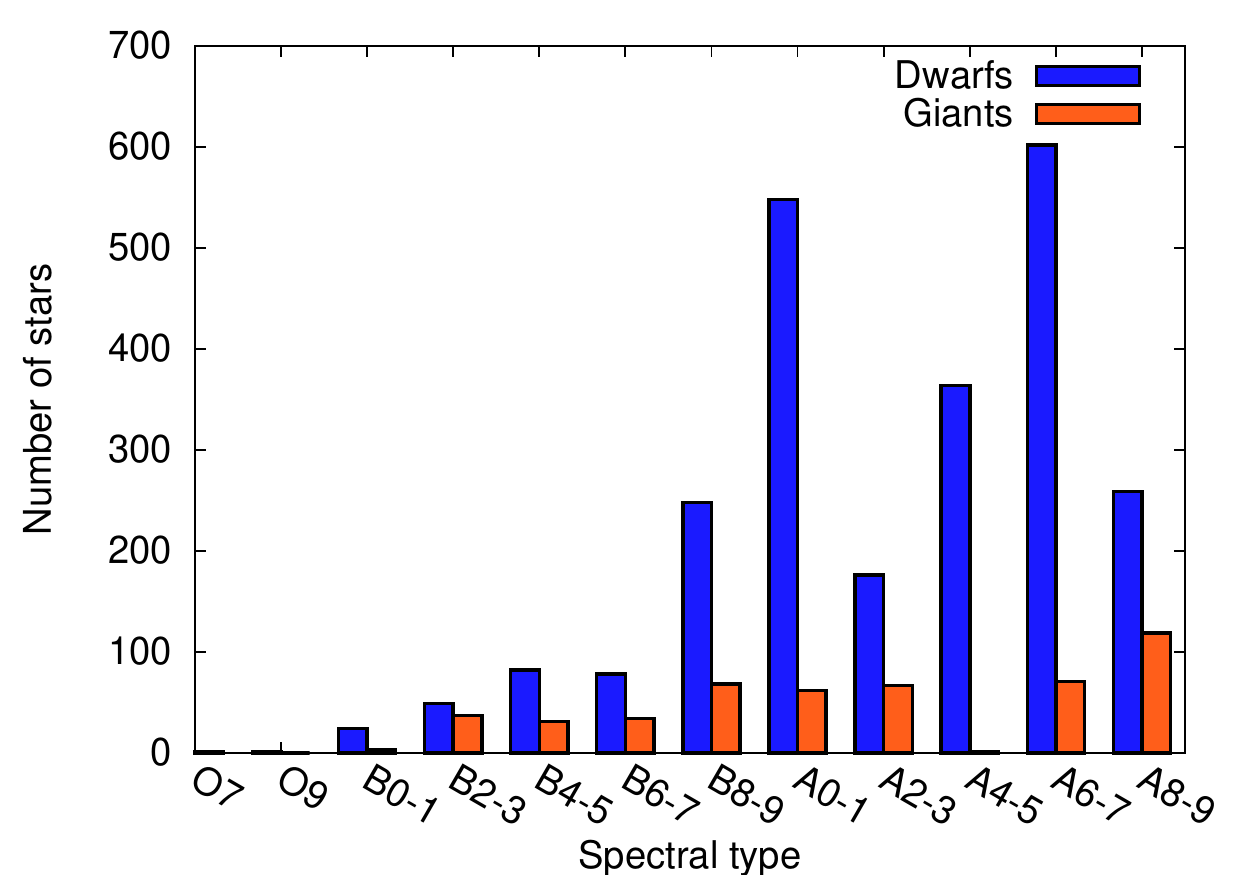}
  \caption{Distribution of spectral types in our sample. dwarf stars and 
sub-giant stars are combined as ``Dwarfs'' while giant stars and super-giant stars are combined to ``Giants''.}
  \label{histogramSpec}
\end{figure}
\section{The spectral types derived}

The full spectroscopic survey contains the spectral types of
11466 stars: 2293 A stars, 655 B stars, two O stars and 
8516 late-type stars (spectral type F to M). The spectral types of all O, B, and A stars are listed in
Table \ref{tab01}.  All these stars are in IRa01, LRa01, 
and LRa02 but not all of them were observed by CoRoT.

The first column in Table \ref{tab01} gives the CoRoT-ID of the stars 
where all 1856 stars actually observed by CoRoT are asterisked.
The second and third columns list the coordinates, and the fourth the brightness of
the star in the V-band. The visual magnitudes were obtained with the
Wide Field Camera filter-system of the Isaac Newton Telescope at Roque de los
Muchachos Observatory on La Palma and can be converted to Landolt
standards \citep{landolt92} as shown in \cite{deleuil09}.  The last column gives the spectral types
which we derived  in this work.

\begin{figure}[h!]
  \includegraphics[height=.26\textheight,angle=0]{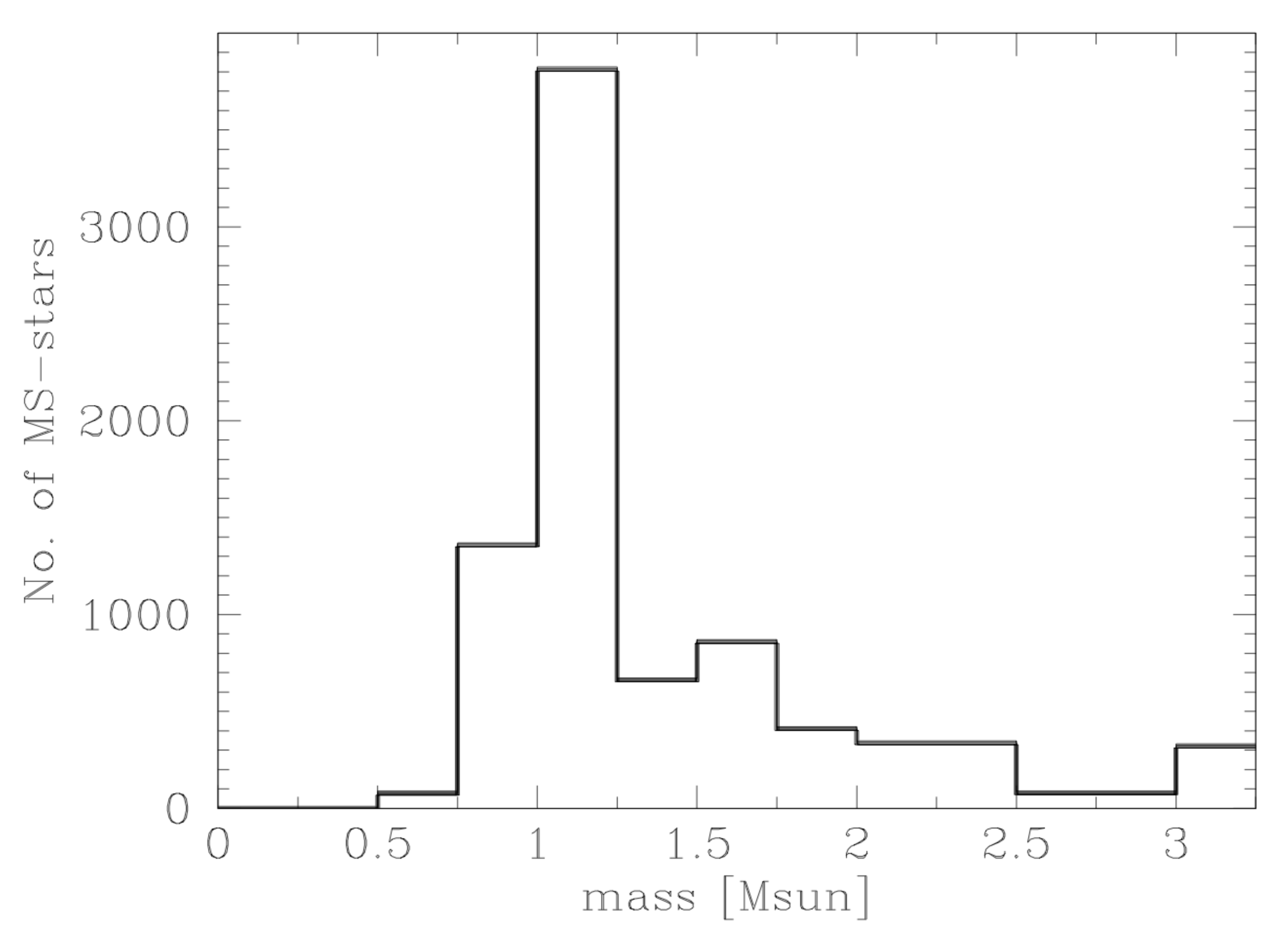}
\caption{Mass function of the main-sequence stars in the three CoRoT fields. 
              The last bin contains the number of stars with masses larger than 3 $M_{\odot}$.}
  \label{histMass}
\end{figure} 

\section{Discussion and Conclusions}\label{phot_compare}

The distribution of spectral types for early-type stars
is shown in Fig. \ref{histogramSpec}. It is interesting to note that
giant stars are rare. $82.5\pm1.7\%$ of the early-type stars are dwarf stars. 
At first glance, the lack of stars with a spectral type
around A3V seems odd.  However, we have to keep in mind that the
temperature- and mass range for this spectral types is much smaller 
than for other spectral types. If we plot the number of stars per mass or
temperature interval, the gap disappears.  Fig. \ref{histMass} shows
the mass-function of main-sequence stars in the IRa01, LRa01, and
LRa02 fields.  Although the mass function peaks around 1.0 $M_{\odot}$, 
the main targets of the CoRoT-survey, there is a long tail towards larger
masses.

In \citet[paper II]{guenther11}, we will show that 
CoRoT detects one planet hosting star for $2100\pm700$ stars observed. 
Using this number, we expect to find one transiting planet
amongst the A, and B stars in IRa01, LRa01, and LRa02.  However,
according to \cite{johnson10a,johnson10b} the
frequency of planets orbiting stars in the mass range between 1.5 and
3 $M_{\odot}$ is a factor of 2 to 4 higher than for stars of 1.0
$M_{\odot}$, which means that we expect to find two to four
transiting planets in our sample. By searching for the phase-dependent
ellipsoidal variations we will even be able to detect massive, close-in
planets and brown dwarfs that are not transiting.

\begin{figure}[h!]
  \includegraphics[height=.25\textheight]{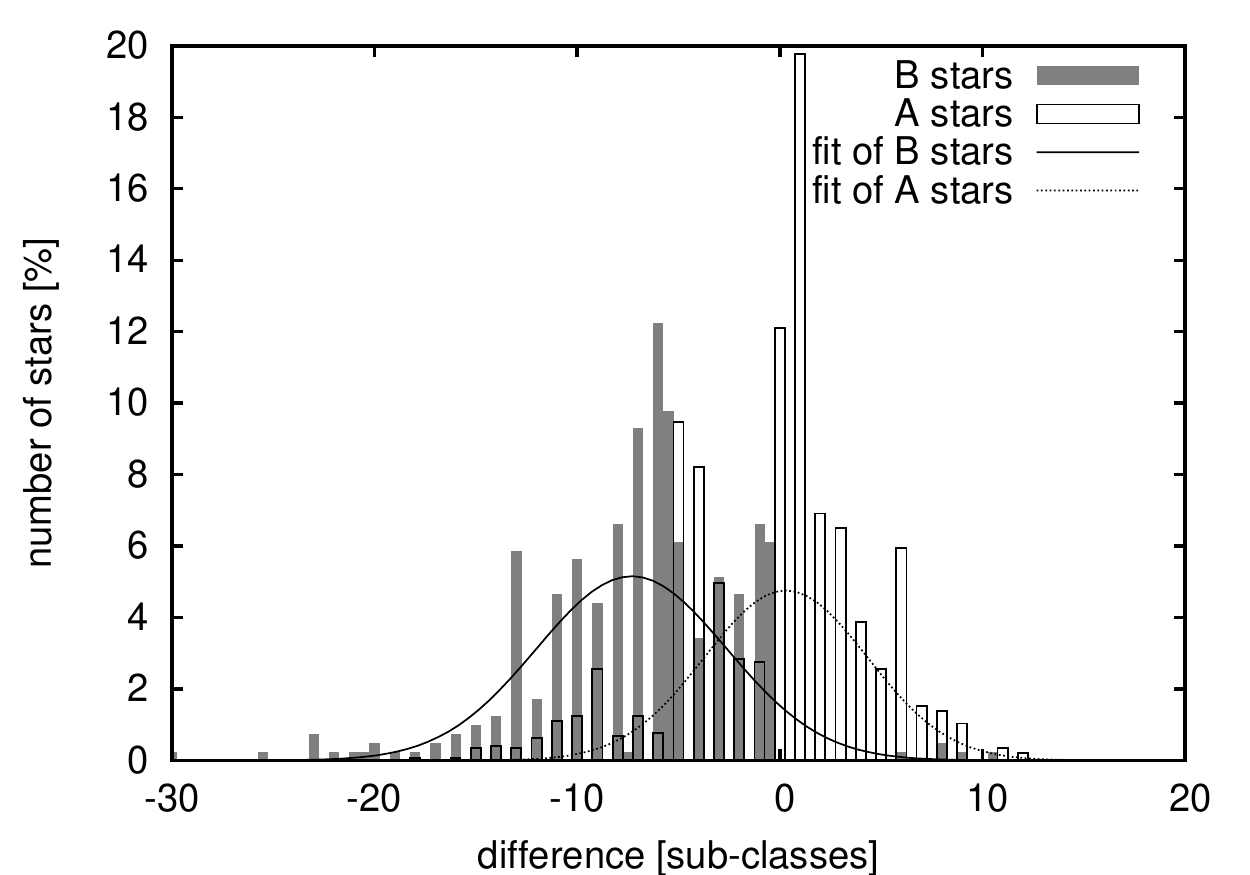}
\caption{Differences between spectroscopic and photometric determination of the spectral types. 
Values smaller than zero indicate that the photometric types are later and values greater than zero 
mean that the photometric types are earlier spectroscopic types. The gaussian fit roughly 
characterizes the distribution of the samples of spectroscopically classified A and B-type stars.}
  \label{SEDspecHist2}
\end{figure}

\begin{figure}[h!]
  \includegraphics[height=.25\textheight,angle=0]{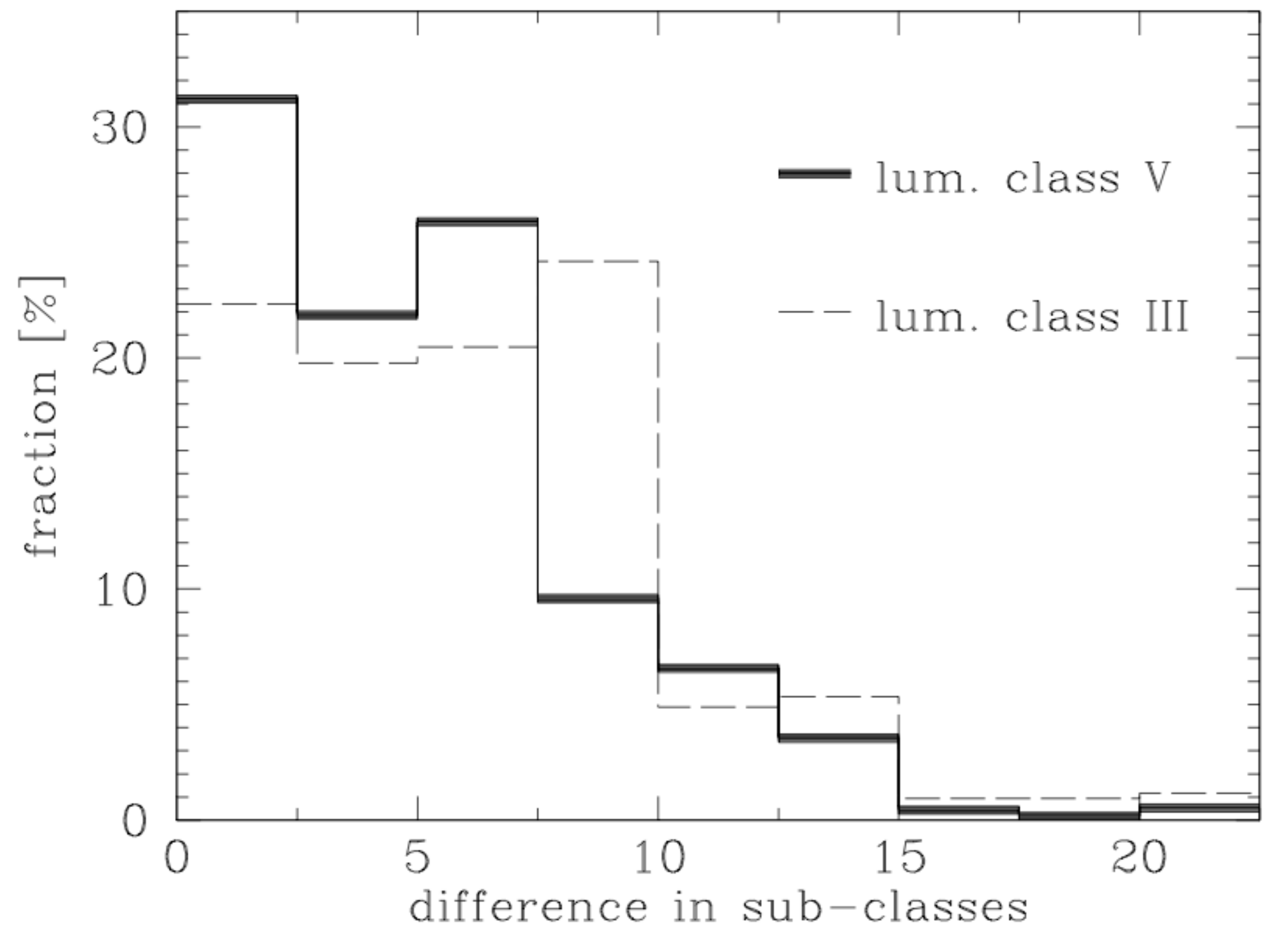}
\caption{Absolute differences between the spectroscopic and photometric 
determination of the spectral types in order to show the difference for 
different luminosity classes.}
  \label{SEDspecHist}
\end{figure}

How well do the spectral-types derived photometrically compare with
those derived spectroscopically?
The photometric classification was taken from the online EXODAT database 
\citep{deleuil09}. It was derived by fitting photometric observations of 
B,V,R,J,H,K$_{\textit{s}}$-colours with spectral-energy-distribution(SED)-templates including 
the reddening. The spectroscopic determination was obtained by fitting 
templates to the observed spectra. Given that the methods used are intrinsically different it is 
not surprising that the results obtained are also different. As explained by \cite{gray09} spectral types are defined by specific spectral features. The analysis of the 
spectra presented here thus is the reference against which the photometric classification has to be 
compared with. Fig. \ref{SEDspecHist2} shows the difference between the photometric and spectroscopic 
classification of the same stars. Negative values represent stars with a photometric type later 
(``lower temperature'') than the spectroscopic one, whereas positive values represent the stars with 
earlier (``higher temperature'') photometric types. 
We note that the difference between the photometric and spectroscopic 
classification is not random but there are systematic differences: 
The photometric classification of B stars always leads to later spectral types 
than the spectroscopic one. In contrast to this A stars peak around identity. 

We interpret this result in the way that the photometric method in general produces statistically correct 
values but the relatively high reddening of the B stars leads to a systematic difference, although the 
extinction was taken into account in the photometric analysis. Strong local reddening in the observed 
CoRoT-fields is also not surprising since there are known star-forming regions like Sh 2-284 
\citep{puga09}.

Moreover there are also stars with extreme strong differences in the classification, often caused by 
blends of more than one star in the fibre. In such a case the integrated light leads to a combined 
spectral type. Intrinsic variability of the stars or unresolved spectroscopic binaries, might also 
be responsible for the different spectral types.

\begin{figure}[h!]
  \includegraphics[height=.35\textheight,angle=270]{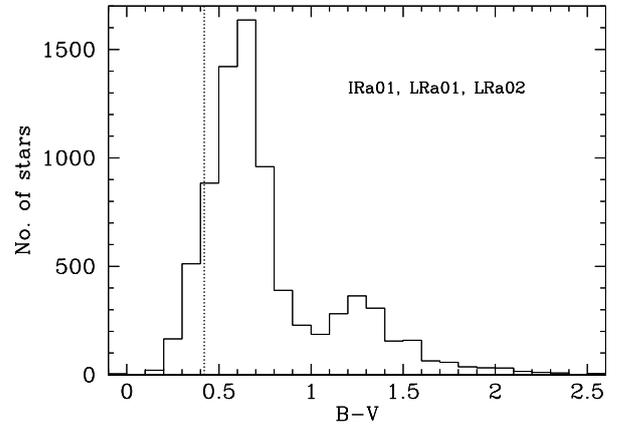}
\caption{B-V colours of stars brighter than $m_{V}=14.5$ in the
IRa01, LRa01, LRa02. 70\% of the stars with $B-V<0 \,\fm42$ (dashed line)
are A  and B stars.}
  \label{histogramPhot}
\end{figure}

Fig. \ref{SEDspecHist} shows the comparison between the photometric spectral types and the
spectral types from this work in absolute values. For 31\% of the stars the spectral types classified 
photometrically and spectroscopically are the same. For more than 50\% of the stars the 
difference is equal or less than five sub-classes. We find that 74\% of the
stars classified photometrically as A and B stars have a spectral type that is earlier than F3.

Because we now know that the contamination by giant stars
is small, and it is possible to identify the A and B stars
photometrically, we can extend the survey to all 19983 early-type stars (A and B stars, observed by 
CoRoT in the long run fields) listed in EXODAT. If we use the $B-V<0\,\fm42$ -criterion 
(corresponding to an un-reddened F5V star, or an A9V
star with A$_V=0.18$ mag \citep{binney98}) 
still 70\% of the preselected stars are real A and B-type stars (see Fig. \ref{histogramPhot}).
The extended sample should lead to the discovery of 10 to 40
planets of early-type stars.

With a sample of planets orbiting intermediate-mass stars we will
test the predictions of planet formation theories, and determine the 
properties of close-in planets orbiting stars with different masses. 

By determining the spectral types of early-type stars in these fields,
it is now not only possible to carry out an efficient survey of
transiting planets orbiting these stars, but it also opens up the
possibility for a study of early-type stars in the CoRoT data-base in
general. 

\begin{acknowledgements}
We are grateful to the user support group of AAT for all their help
and assistance for preparing and carrying out the observations.  We
would like to particularly thank Rob Sharp, Fred Watson and Quentin
Parker. The authors thank DLR and the German BMBF for the support
under grants 50 OW 0204, and 50 OW 0603 as well as DFG for the support
of S. Geier through grant HE1356/49-1.
\end{acknowledgements}

\onecolumn
\begin{center}
\begin{longtable}{l l r l c c}
\caption{Early-type stars in IRa01, LRa01, and LRa02} \label{tab01} \\

\hline
CoRoT-id & RA [deg]      &DEC [deg]      & V [mag]   & Spec \\
         &(2000.0) &(2000.0) && AAOmega \\
\hline
\endfirsthead

\multicolumn{4}{c}%
{{ \tablename\ \thetable{} -- continued}} \\
\hline 
CoRoT-id & RA [deg]      &DEC [deg]      & V [mag]   & Spec \\
         &(2000.0) &(2000.0) && AAOmega \\
\hline
\endhead

\endfoot

\hline \hline
\endlastfoot
102777938&101.35083&0.22969&11.473&O7III\\
103074452&103.14964&-2.95526&12.755&O9V\\
102807548$^{*}$&101.550789&-2.91421&12.479&B0.5Ia\\
102728404$^{*}$&101.072823&0.34213&11.928&B0IV\\
102785204$^{*}$&101.390648&0.33614&12.058&B0IV\\
102580983$^{*}$&100.22567&-0.61863&12.913&B0IV\\
102593959$^{*}$&100.318832&-1.12783&13.479&B0IV\\
102745264&101.167007&0.25597&10.79&B1II\\
\end{longtable}
\end{center}
\twocolumn

\bibliographystyle{aa}
\bibliography{general}

\end{document}